# Static Spherically Symmetric Solution of $(R \pm \mu^4/R)$ Gravity

$^{a,b}$**Kh. Saaidi** [1], $^{a}$**A. Vajdi** [2] and $^{b,c}$**A. Aghamohammadi**[3]

$^a$*Department of Physics, Faculty of Science, University of Kurdistan, Pasdaran Ave., Sanandaj, Iran*
$^b$*Islamic Azad University of snandaj, Sanandaj, Iran*
$^c$*Plasma Physics Research Center, Science and Research Branch Islamic Azad University of Tehran, Iran*

**Abstract**

The static spherically symmetric solution for $R \pm \mu^4/R$ model of $f(R)$ gravity is investigated. We obtain the metric for space-time in the solar system that reduces to the Schwarzschild metric, when $\mu$ tends to zero. For the obtained metric, the deviation from Einstein gravity is very small. This result is different from the other results have been obtained by equivalence between $f(R)$ gravity and scalar tensor theory. Also it is shown that the vacuum solution in the solar system depends on the shape of matter distribution which differ from the Einstein's gravity.

---

[1] E-mail: ksaaidi@uok.ac.ir
[2] E-mail: Avajdi@uok.ac.ir
[3] Email:agha35484@yahoo.com

# 1 Introductions

Observations on supernova type Ia [1], cosmic microwave background [2] and large scale structure [3], all indicate that the expansion of the universe is not proceeding as predicted by general relativity, if the universe is homogeneous, spatially flat, and filled with relativistic matter. This problem led to several theoretical models for example quintessential scenarios [4] which generalize the cosmological constant approach [5], higher dimensional scenarios [6, 7] or the resort to cosmological fluids with exotic equation of state [8]. Also, an interesting approach is $f(R)$ theories of gravity which generalize the geometrical part of Hilbert-Einstein lagrangian [10, 11, 13, 14, 16]. Indeed it was shown [17] it is possible to construct $f(R)$ model realizing any given cosmology. One of the initiative $f(R)$ model supposed to explaining the positive acceleration of expanding universe has $f(R)$ action as $f(R) = R - \mu^4/R$ [11]. In this model, for large values of Ricci scalar, $R \gg \mu^2$, $f(R)$ function tends to $f(R) = R$, so we expect for these values of $R$ the modification become negligible but for small values of Ricci scalar, $R \lesssim \mu^2$, $f(R)$ action is not the linear one thus for this values of Ricci scalar gravity is modified. Because for large values of $R$ the modification is negligible, this model can not explain the inflation but there is several viable models unifying inflation and late time acceleration[18].

After proposing the $f(R) = R - \mu^4/R$ model it was appeared this model suffer several problems. In the metric formalism, initially Dolgov an Kawasaki discovered the violent instability in the matter sector [19]. The analysis of this instability generalized to arbitrary $f(R)$ models [20, 21] and it was shown an $f(R)$ model is stable if $d^2f/dR^2 > 0$ and unstable if $d^2f/dR^2 < 0$. Thus we can deduce $R - \mu^4/R$ suffer the Dolgov-Kawasaki instability but this instability removes in the $R + \mu^4/R$ model, where $\mu^4 > 0$. Of course, it was shown adding the positive power of Ricci scalar to $R - \mu^4/R$ model removes this instability [14] and later, a number of models which don't suffer this instability appeared in [18]. Furthermore one can see in the $R - \mu^4/R$ model the cosmology is inconsistent with observation when non-relativistic matter is present. In fact there is no matter dominant era [22, 23]. However the recent study shows the standard epoch of matter domination can be obtained in the $R + \mu^4/R$ model [24].

It is obvious that a viable theory of gravity must have the correct newtonian limit. Indeed a viable theory of $f(R)$ gravity must pass solar system tests. After the $R - \mu^4/R$ was suggested as the solution of cosmic-acceleration puzzle [11], Chiba [25] argued that this theory is inconsistent with solar system tests. This claim was based on the fact that metric $f(R)$ gravity is equivalent to $\omega = 0$ Brans-Dicke theory, while the observational constraint is $\omega > 40000$. In fact Chiba claim that if modifications become important recently the scalar field is generically light so the value of post newtonian parameter $\gamma_{PPN}$ is practically independent of the mass of scalar field. Thus in this theory $\gamma_{PPN} = 1/2$. But this is not quite the case and by using equivalence between $f(R)$ gravity and scalar tensor gravity one can find models which are consistent with the solar system tests where this consistent can be made by giving the scalar a high mass or by exploiting the so-called chameleon effect[26]. If the mass of scalar field is large the scalar become short-ranged and therefore has no effect in solar system test. Typical models with high mass have been considered in [25, 14]. In [14] the model with positive and negative power of Ricci scalar has been proposed and the present universe has been considered as a deSitter space with the large scalar mass. In the chameleon fields the scalar degree of freedom acquires a large effective mass at solar system scales while the effective mass is light at cosmological scales [27, 28, 15]. Indeed for these models the effective mass of scalar degree of freedom depends on the density of matter. Furthermore it is possible to investigate the spherical symmetric solutions of $f(R)$ gravity with out invoking the equivalence of $f(R)$ gravity and scalar tensor theory [29, 30, 31, 32, 33, 34]. It was shown that some $f(R)$ models accept the Schwarzschild-de Sitter spacetime as a spherical symmetric solutions of field equation[32, 29, 12]. Hence $R - \mu^4/R$ model has a Schwarzschild-de Sitter solution with constant curvature as $R = \sqrt{3\mu^4}$ unlike the $R + \mu^4/R$ model. However acceptances of Schwarzschild-de Sitter spacetime as a metric at solar system has been contested [33, 34]. In [33] the Newtonian limit of $R - \mu^4/R$ model has been studied by linearized filed equations. It is important to note that in [33], authors use an approximation, which $R$ is continuous and in order of $\mu^2$. They have obtained a metric for space-time in the solar system which does not reduce to general relativity in the $\mu \to 0$ limit, because, their assumptions are not consistent with ordinary general relativity.

In this paper, we investigate the static spherically symmetric solution of $R \pm \mu^4/R$ model of $f(R)$ gravity without invoking scalar tensor theory. We have not used the assumptions of [32] and we considered this model with continuity constraint on the components of metric. We determine a metric for space-time in the solar system which reduces to general relativity in the $\mu \to 0$ limit which is not the case in [33, 34], and we find out that in the solar system the deviation from general relativity is very small.

## 2 Vacuum solution in the solar system

We consider modified gravitational theory with an action of the form

$$S = \frac{1}{16\pi G}\int \sqrt{-g}\left[R \pm \frac{\mu^4}{R}\right]d^4x + S_m, \tag{1}$$

where $R$ is Ricci scalar and $S_m$ is the matter action. The field equation resulting from this action in the metric approach are

$$\left(1\mp \frac{\mu^4}{R^2}\right)R_{\mu\nu} - \frac{1}{2}\left(1\pm \frac{\mu^4}{R^2}\right)Rg_{\mu\nu} \mp \mu^4\left(\nabla_\mu\nabla_\nu - g_{\mu\nu}\nabla_\alpha\nabla^\alpha\right)\frac{1}{R^2} = 8\pi GT_{\mu\nu}. \tag{2}$$

The Newtonian limit of $R - \mu^4/R$ model has been studied in [33], which the Newtonian limit of gravitational field in the solar system for outside of matter has been obtained as

$$ds^2 = -\left(1 - \frac{2M}{r}\right)dt^2 + \left(1 + \frac{M}{r}\right)dr^2 + r^2 d\Omega^2, \tag{3}$$

It is clearly seen that $\gamma_{PPN} = \frac{1}{2}$ for the metric (3), while the measurements give $\gamma_{PPN} = 1 + (2.1 \pm 2.3) \times 10^{-5}$ [9]. We note that for $\mu = 0$ equations (1) and (2) reduce to Hilbert-Einstein action and Einstein's equation respectively, but the metric (3), which has been obtained as a solution of the field equation (2), does not reduce to Schwarzschild metric. The origin of this matter is:
i) The authors of [33] use an approximation which is based on the equivalence between the $f(R)$ gravity and Brans-Dicke theory with $\omega = 0$.
ii) They have obtained $R \sim \mathcal{O}(\mu^2)$. In this case, order of deviation from general relativity is approximately 1, i.e., $\mu^4/R^2 \sim \mathcal{O}(1)$, which is independent of $\mu$. Therefore in the limit $\mu \to 0$, the deviation from general relativity do not tend to zero. Since the modified field equation (2) reduces to Einstein's equation when $\mu$ tends to zero, then we expect that the metric and space curvature of this model reduces to the metric and space curvature which has been obtain from Einstein's field equation. By the way, we know that the Ricci scalar of Einstein's gravity is $R = -8\pi GT/c^4$, where $T$ is the trace of the energy momentum tensor. So, we assume that the space curvature of the modified field equation (2) is as a perturbation of space curvature of Einstein's gravity

$$R = -8\pi GT + R_p, \tag{4}$$

where $R_p$ is a perturbation term which comes from action modification. By this point of view, we have obtained another solution for $(R \pm \mu^4/R)$ model of $f(R)$ theory in the solar system. It is well known that the static spherically symmetric solution for Einstein's action is Schwarzschild metric. Since $\mu$ is very small, and $\mu$ tends to zero, the action (1) reduces to Einstein's action, hence we assume that the action (1) is a perturbation of Einstein's action. Therefore we can treat the space time as a small perturbation to the Schwarzschild space time as follows

$$ds^2 = -\left(1 - \frac{2M}{r} + a(r)\right) dt^2 + \left(1 - \frac{2M}{r} + b(r)\right)^{-1} dr^2 + r^2 d\Omega^2. \quad (5)$$

One can obtain the Ricci scalar and Einstein's tensor as

$$R = -2\left(\frac{b}{r^2} + \frac{b'}{r} + \frac{a''}{2} + \frac{a'}{r}\right), \quad (6)$$

$$\begin{aligned}
(G)^r_r &= \frac{b}{r^2} + \frac{a'}{r}, \\
(G)^\theta_\theta &= (G)^\varphi_\varphi = \frac{b'}{2r} + \frac{a''}{2} + \frac{a'}{2r}, \\
(G)^t_t &= \frac{b}{r^2} + \frac{b'}{r}.
\end{aligned} \quad (7)$$

where,($\prime$) denotes the derivation with respect to $r$. From equations (6) and (7), it is clear that the first non zero order of Ricci Scalar and Einstein's tensor are only a function of perturbation terms $a(r)$ and $b(r)$. This result is agreement with those which obtain from Schwarzschild metric. In empty space we rearrange the field equation (2) as the following

$$R^\nu_\mu = (1 + \phi)^{-1} \left[\frac{R}{2}(1 - \phi)\delta^\nu_\mu - (\delta^\nu_\mu \Box - \nabla_\mu \nabla^\nu)\phi\right], \quad (8)$$

where $\phi = \mp \frac{\mu^4}{R^2}$. Expanding $(1 - \phi)^{-1}$ and keeping the terms of equation (8) up to first order of $\phi$, we have

$$(G)^\nu_\mu = (R)^\nu_\mu - \frac{(R)}{2}\delta^\nu_\mu = \left(\nabla_\mu \nabla^\nu - \delta^\nu_\mu \Box\right) \phi. \quad (9)$$

From (6) it is clearly seen that, $R$ is a function of perturbation parameter only, in which by limiting $\mu \to 0$ is very small. Therefore for obtaining (9) we have neglected $R\phi$. By calculating the right hand of (9) up to the first order, we obtain

$$\begin{aligned}
\phi_{;r}^{;r} &= \phi'', \\
\phi_{;\theta}^{;\theta} &= \phi_{;\varphi}^{;\varphi} = \frac{\phi'}{r}, \\
\phi_{;t}^{;t} &= \frac{\phi' a'}{2}, \\
\Box\phi &= \nabla^2 \phi = \phi'' + \frac{2\phi'}{r}.
\end{aligned} \qquad (10)$$

By substituting the set of equations (7) and (10) in (9), one can arrive at

$$\begin{aligned}
\frac{b}{r} + a' &= -2\phi', \\
b' + ra'' + a' &= -2(r\phi')', \\
b + rb' &= -(r^2 \phi')'.
\end{aligned} \qquad (11)$$

Using (6) in $\phi = \mp \mu^4/R^2$ and then by substituting it in the set of equations (11) we find expressions for $a(r), b(r), R(r), \phi(r)$ as

$$\begin{aligned}
b &= \pm \alpha (\mu r)^{\frac{4}{3}}, \\
a &= \pm \frac{3}{4} \alpha (\mu r)^{\frac{4}{3}}, \\
R &= \mp 7 \alpha \mu^{\frac{4}{3}} r^{-\frac{2}{3}}, \\
\phi &= \mp \frac{1}{49\alpha^2} (\mu r)^{\frac{4}{3}},
\end{aligned} \qquad (12)$$

where $\alpha^3 = 4/147$. It is obvious that, (12), in the limit $\mu \to 0$ the Ricci scalar and the deviation of general relativity, tend to zero. This result is agreement with what has been obtained from the Schwarzschild metric. Therefore we can approximate the static spherically symmetric metric of $R \pm \mu^4/R^2$ model for empty space, $T_{\mu\nu} = 0$ in the solar system, as

$$\begin{aligned}
ds^2 = &- \left[ 1 - \frac{2M}{r} \pm \frac{3}{4} \alpha (\mu r)^{\frac{4}{3}} \right] dt^2 + \left[ 1 - \frac{2M}{r} \pm \alpha (\mu r)^{\frac{4}{3}} \right]^{-1} dr^2 \\
&+ r^2 d\Omega^2.
\end{aligned} \qquad (13)$$

It is clear that the above metric reduces to Schwarzschild metric in case $\mu = 0$. And this subject shows that this solution is in agreement with field equation (2) which for $\mu = 0$ reduces to Einstein's equation.

## 3 Interior solution in the solar system

We now consider the metric for space-time inside of matter for the solar system. Inside matter, field equation (2) can written as

$$R_\mu^\nu = (1+\phi)^{-1}\left[\frac{R}{2}(1-\phi)\delta_\mu^\nu - (\delta_\mu^\nu \Box - \nabla_\mu \nabla^\nu)\phi + \frac{8\pi G}{c^4}T_\mu^\nu\right], \qquad (14)$$

where

$$\frac{8\pi G}{c^4}T = \mathcal{O}\left(\frac{M}{r_0^3}\right), \qquad (15)$$

where $r_0$ is the radius of mater distribution. Because of existence the term $(8\pi G/c^4)T_{\mu\nu}$ in right side of of field equation (14) the structure of space-time inside matter is different from vacuum case. We note that in a typical solar system we have $M/r_0 \gg (\mu r_0)^{4/3}$, which leads

$$\frac{(\mu r_0)^4}{(\frac{M}{r_0})^2} \ll (\mu r_0)^{\frac{4}{3}} \ll \frac{M}{r_0}. \qquad (16)$$

From numerical point of view one can investigate (16). According to cosmological studies, we have $\mu^2 = 10^{-52}m^{-2}$ [11], and, as example, for Sun, $r_0 = 6.9 \times 10^8 m$ and Schwarzschild radius is $2M = 2.96 \times 10^3 m$, which results in

$$(\mu r_0)^{\frac{4}{3}} \sim \left(\frac{M}{r_0}\right)^4,$$
$$\frac{(\mu r_0)^4}{(\frac{M}{r_0})^2} \sim (\frac{M}{r_0})^{10}. \qquad (17)$$

According to equations (16), (17) it is obvious that for $R \sim \mathcal{O}(M/r_0^3)$, the order of deviation term is

$$(\frac{\mu^4}{R^2}) = \frac{(\mu r_0)^4}{(\frac{M}{r_0})^2} \ll (\mu r_0)^{\frac{4}{3}} \ll \frac{M}{r_0}. \qquad (18)$$

So inside matter distribution with assuming $R \sim \mathcal{O}(M/r_0^3)$, deviation from general relativity is smaller than $(\mu r_0)^{4/3}$ and at least up to the order $(\mu r_0)^{4/3}$ the solution of field equation (14) is the same as that of Einstein's equation. Hence, up to the order $(\mu r_0)^{4/3}$, we can use the solution of Einstein's equation as the solution of field equation (14). The solution of Einstein's equation inside matter is [36]

$$\begin{aligned}ds^2 &= -e^{\nu(r)}dt^2 + e^{\lambda(r)}dr^2 + r^2 d\Omega^2, \\ &= -e^{\nu(r)}dt^2 + \left[1 - \frac{8\pi G}{r}\int_0^r \rho_m(y)y^2 dy\right]^{-1}dr^2 + r^2 d\Omega^2,\end{aligned} \qquad (19)$$

where $e^\nu$ is the solution of equations

$$-\frac{1}{2}\frac{d\nu}{dr}(p+\rho_m) = \frac{dp}{dr}$$
$$8\pi G(p+\rho_m) = \frac{e^{-\lambda}}{r}\frac{d(\nu+\lambda)}{dr}. \quad (20)$$

Continuity of metric (13,19) at the boundary of matter distribution ($r = r_0$), results in

$$\frac{8\pi G}{r_0}\int_0^{r_0}\rho_m(y)y^2 dy = \frac{2M}{r_0} \mp \alpha(\mu r_0)^{\frac{4}{3}}. \quad (21)$$

From equation (21) it is obvious the physical interpretation of the parameter ($M$) is not the same as the case general relativity. The ordinary relation

$$4\pi G\int_0^{r_0}\rho_m(y)y^2 dy = M, \quad (22)$$

does not hold in $f(R)$ gravity. The relation between total mass ($m$) and parameter ($M$) in $R \pm \mu^4/R$ gravity is

$$Gm \pm \alpha\mu^{\frac{4}{3}}r_0^{\frac{7}{3}} = M,$$
$$4\pi\int_0^{r_0}\rho_m(y)y^2 dy = m. \quad (23)$$

The above equation demonstrates that the external solution depends on the shape of matter distribution which is different from general relativity.

## 4 Gravitational redshift in $R \pm \mu^4/R$ gravity

Frequency $\nu_1$ and $\nu_2$ measured by observers at points $P_1$ and $P_2$ are related by [36]

$$\frac{\nu_1}{\nu_2} = \sqrt{\frac{g_{tt}(2)}{g_{tt}(1)}}. \quad (24)$$

For metric (13) the above equation reduces to

$$\frac{\nu_1}{\nu_2} \simeq \sqrt{\frac{1-\frac{2M}{r_2}}{1-\frac{2M}{r_1}}}\left[1 \mp \frac{3}{8}\alpha\mu^{\frac{4}{3}}(r_1^{4/3} - r_2^{4/3})\right]. \quad (25)$$

If we assume a clock comparison between Earth and a satellite at $15 \times 10^6 m$ height, equation (25) leads to deviation from general relativity at the order of

$$\frac{3}{8}\alpha\mu^{\frac{4}{3}}(r_1^{4/3} - r_2^{4/3}) \sim \mathcal{O}(10^{-25}). \tag{26}$$

Clock comparison can perform with an accuracy of $\mathcal{O}(10^{-15})$ [35], so we conclude the deviation from general relativity for this test is smaller than the accuracy of experiment.

## 5  Discussion

We have investigated the static spherically symmetric solution for $R \pm \mu^4/R$ model of $f(R)$ gravity. We have determined the metric for space-time in the solar system. We have shown $R \pm \mu^4/R$ model of $f(R)$ gravity reduces to the general relativity in the $\mu \to 0$ limit. It is seen that, the metric which has been found, differ slightly form general relativity, so it's predictions in solar system tests is approximately similar to general relativity results. Also we have shown that the vacuum solution in the solar system depends upon the shape of matter distribution which is different from general relativity. The metric which we have obtained is complectly different from what has been obtained in [33]. They supposed that the space time is a small perturbation of the de Sitter space time and solved the trace field equation for this model at the case which Ricci scalar is continuous. Then, they used continuous constraint on $R$ and found the space time metric for the model. It is well known that this emphasis (the continuity of Ricci Scalar ) is not consistent with ordinary general relativity theory. Therefore, the result of their consideration is different with general relativity theory. Although, we have restricted our analysis to $R \pm \mu^4/R$ model, similar result can be derived for other $f(R)$ model, such as $f(R) = R + \epsilon h(R)$ in which when $\epsilon$ tends to zero the Ricci scalar reduces to zero.